\documentclass[prb,aps,twocolumn,epsfig,eqsecnum]{revtex4}

\newcommand{\beq}{\begin{equation}}
\newcommand{\eeq}{\end{equation}}
\newcommand{\bea}{\begin{eqnarray}}
\newcommand{\eea}{\end{eqnarray}}

\newcommand{\etal}{{\em et al.}}

\newcommand{\Tr}{{\rm Tr}}

\newcommand{\cD}{{\cal D}}
\newcommand{\cG}{{\cal G}}
\newcommand{\cF}{{\cal F}}
\newcommand{\ij}{_{ij}}
\newcommand{\Pf}{{\mathrm{Pf}}}
\newcommand{\imag}{i}

\def\tit#1#2#3#4#5{{#1}{\bf #2}, #3 (#4)}
\def\jmp{J.\ Math.\ Phys.\ }

\def\npb{Nucl.\ Phys.\ B\ }

\def\prl{Phys.\ Rev.\ Lett.\ }
\def\pr{Phys.\ Rev.\ }
\def\prb{Phys.\ Rev.\ B\ }

\def\zpb{Z.\ Phys.\ B\ }

\usepackage{epsfig}
\usepackage{wasysym}

\begin{document}


\title{On Ising and dimer models in two and three dimensions}

\author{R. Moessner$^1$ and S. L. Sondhi$^2$}

\affiliation{$^1$Laboratoire de Physique Th\'eorique de l'Ecole Normale
Sup\'erieure, CNRS-UMR8541, Paris, France}
\affiliation{$^2$Department of Physics, Princeton University,
Princeton, NJ 08544, USA}

\date{\today}

\begin{abstract}
Motivated by recent interest in 2+1 dimensional quantum dimer models,
we revisit Fisher's mapping of two dimensional Ising models 
to hardcore dimer models. First, we note that the symmetry breaking
transition of the ferromagetic Ising model maps onto a non-symmetry 
breaking transition in dimer language---instead it becomes a 
deconfinement transition for test monomers.
Next, we introduce a modification of Fisher's mapping in which a
second dimer model, also equivalent to the Ising model, is defined on a 
generically different lattice derived from the dual. In contrast to Fisher's 
original mapping, this enables us to reformulate frustrated Ising models as dimer 
models with positive weights and we illustrate this by providing a new 
solution of the fully frustrated Ising model on the square lattice. 
Finally, by means of the modified mapping we show that a large class of 
three-dimensional Ising models are precisely equivalent, in the time
continuum limit, to particular quantum dimer models. 
As Ising models in three 
dimensions are dual to Ising gauge theories, this further yields an exact map 
between the latter and the quantum dimer models.
The paramagnetic phase in Ising language maps onto a deconfined, 
topologically 
ordered phase in the dimer models. Using this set of ideas, we also
construct an exactly soluble quantum eight vertex model.
\end{abstract}



\maketitle

\section{Introduction}
\begin{figure}
\centerline{ \psfig{figure=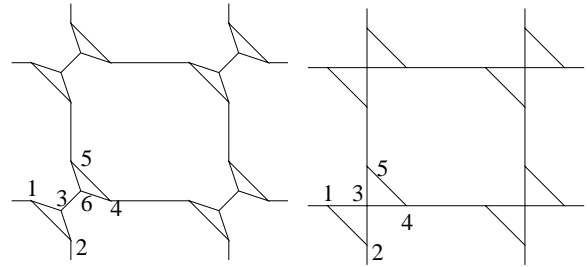,height=3.5cm} }
\caption{
Left: the Fisher lattice, a decorated square lattice. The six sites in
the basis correspond to one square lattice site. A dimer model
on this lattice can be used to calculate the partition function of
the ferromagnetic Ising model on the (direct) square lattice. Right:
The frustrated Fisher lattice, suitable for a dimer model representing the
Ising model on the fully frustrated (dual) square lattice.}
\label{fig:fishlat}
\end{figure}

Dimer models have long been of interest to statistical 
mechanicians.\cite{rushfow,kasteleyn,nagle,gaunt,liebhei} 
In addition to their interest in various physical
contexts, they have the striking feature that they
are exactly soluble on any planar graph.\cite{kasteleyn} 
Following this insight, Fisher,\cite{FishIsing} after
initial work by Stephenson,\cite{stephensonising} 
constructed a general mapping -- reviewed
below -- from two-dimensional Ising onto dimer models, thereby
relating the solvability of the one to that of the other. 
In particular he related the partition function of the
ferromagnetic Ising model on the square lattice to that
of the dimer model on the (now) Fisher lattice, which is
sketched in Fig.~\ref{fig:fishlat}.

More recently, {\it quantum} dimer models (QDMs) have been formulated
and studied.\cite{Rokhsar88}  These live in Hilbert spaces spanned by 
dimer configurations of a given lattice and their Hamiltonians
contain kinetic and potential energies that are naturally
defined in this basis. These models were introduced  to capture
the dynamics of valence bond dominated phases of quantum antiferromagnets,
with the particular intent of finding Anderson's hypothesized resonating 
valence bond (RVB) liquid\cite{pwabook} -- 
a hope realized recently.\cite{MStrirvb}

In this context, attention has been focussed on the nature of the
quantum dimer phases, in particular their topological properties and
low-energy excitations, and a recurring and useful theme has been their
interpretation in gauge theoretic terms, particularly the identification
of the RVB phase as a deconfined 
phase.\cite{eduardosteve,wen,sentfish,msf,misserpas} 
In Ref.~\onlinecite{msf} this identification was made as an exact reduction of an
``odd'' Ising gauge theory to a quantum dimer model on the same lattice,
in the extreme strong coupling limit. As Ising gauge theories in $d=2+1$ are dual
to Ising models, this allows dimer models to be exactly related to frustrated
quantum (transverse field) Ising models in $d=2+1$ in the limit of weak
transverse fields and thence to a set of ideas and techniques for
obtaining their phase diagrams.\cite{msc}
Very recently, Misguich \etal\ have constructed an exactly soluble
dimer model on the kagome lattice.  This model maps onto a transverse
field Ising model with {\em zero} exchange, thereby allowing the
entire spectrum and correlations to be determined.\cite{misserpas}

In this note, we further explore the above connections between Ising
and dimer models in two and three dimensions. We begin by reviewing
Fisher's construction in two dimensions which utilises the loop model 
generated by the Ising high-temperature expansion, which in turn is 
mapped onto a
dimer model on a decorated lattice. We note that the symmetry breaking 
transition in the two dimensional ferromagnetic Ising model maps onto
a non-symmetry breaking, deconfinement transition of test monomers in the 
dimer model. As an aside we point out
that the dimer formulation provides an
immediate insight into how the Ising model yields a lattice theory
without doubled fermions. We next observe that in the presence of
frustration in the Ising model, Fisher's construction leads to negative 
weights 
in the dimer model. To remedy this, we introduce a modification of
Fisher's construction which proceeds via an intermediate map to 
a generalised (non-hardcore) dimer model on the dual lattice which is then
decorated to produce the hardcore constraint. As an example,
we solve the fully frustrated Ising model on the square lattice as a
dimer model on a modified Fisher lattice.

\begin{figure}
\centerline{ \psfig{figure=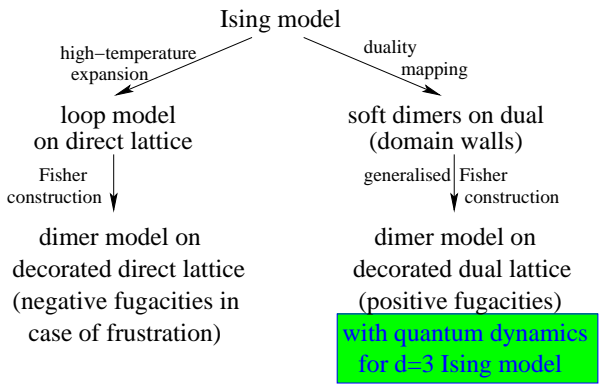,height=5cm} }
\caption{ 
The mappings of the Ising model onto dimer models. 
}
\label{fig:mapofmaps}
\end{figure}

Finally, we turn to a class of
three-dimensional classical Ising models---these are 
general (frustrated or unfrustrated) two-dimensional Ising magnets
stacked ferro- or antiferromagnetically in the third direction,
which can therefore be mapped onto transverse field Ising magnets in
2+1 dimensions by taking a continuum limit in the third direction. 
Armed with the modified Fisher construction we are able to map
these onto quantum dimer models where the quantum dynamics induced by 
the transverse field translates to a particular ``resonance'' dynamics
of the kind studied by Misguich \etal\ This is particularly interesting
from the viewpoint of the dimer models since it allows known lore
on Ising models to be transcribed into statements about the former.
Quite generally, the paramagnetic phases of the Ising models map
to deconfined, topologically ordered phases in the dimer models.
Specifically, for the case of the ferromagnetic Ising model we find
a dimer transition characterized solely by confinement and the loss
of topological order. We also observe that the dimer models 
are exactly equivalent to Ising gauge theories, as they must be since
the latter are dual to Ising models in 2+1 dimensions. Consequently
their spectrum in the deconfined phase is understood in terms of
Ising vortices or visons. The relationships between these mappings
are displayed in Fig.~\ref{fig:mapofmaps}.

We close with some remarks concerning possible extensions of this
work, including the construction of an exactly soluble quantum
eight-vertex model along the lines of Ref.~\onlinecite{misserpas}.
Such a model may also be of interest in the context of orbital current
models, such as the d-density wave models considered in
Ref.~\onlinecite{sudipvert}.

\section{The Ising transition in $d=2$ as a deconfinement transition}
In this and further sections, we consider classical Ising models
defined by the Hamiltonian
\bea
H=-\sum_{\left<ij\right>}J_{ij}\sigma_i^z \sigma_j^z,
\label{eq:classham}
\eea
where $\sigma^z$ is a Pauli matrix, and $J_{ij}$ is the strength of
bond $\left<ij\right>$. The sites of the lattice, labelled by $i=1
\cdots N$, together with the bonds $\left<ij\right>$ with nonzero
$J_{ij}$ form a graph, ${\cal G}$, the interaction graph. In the
following, we will use the following definitions: $\beta=1/k_B T$;
$K_{ij}=\beta J_{ij}$; $v\ij=1/w\ij=\tanh K_{ij}$. Although we have
formally written a quantum Hamiltonian, the problem is, of course,
still classical.

We first briefly review Fisher's mapping.\cite{FishIsing} 
Its starting point is the
high temperature expansion of the partition function, 
$Z\Tr\exp(-\beta H)$, of the
Ising model: 
\bea
Z=2^N 
\left[
\prod_{\left<ij\right>} \exp(-K_{ij}) \cosh(K_{ij}) 
\right]
\Upsilon(v\ij;\cG).
\eea Here, $\Upsilon(v\ij;\cG)$ is the crucial quantity: it is
the sum over all loop coverings, labelled by $\Gamma(\cG)$, of the
graph $\cG$. These loops can intersect one another, provided an even
number of links emanate from each site. The weighting of a particular
covering is given by the product of $v\ij$ of all the links of its
loops, so that 
\bea
\Upsilon(v\ij;\cG)=\sum_{\Gamma(\cG)}
\prod_{\left<kl\right>\in\cG} v_{kl}\eea\ .

Fisher's mapping turns the resulting loop model on the graph $\cG$
into a dimer model on a decorated graph, $\cD$, by a decoration
procedure outlined in Ref.~\onlinecite{FishIsing}, see in particular
its Fig.~6 for the general decoration rule and Fig.~7 for the explicit
example of the square lattice magnet. The resulting lattice is
depcited in Fig.~\ref{fig:fishlat}. There are now two types of bonds,
the original (`external') ones and the new (`interior') ones.

The crucial property of Fisher's mapping is that, if $\cG$ is a planar
graph, so is $\cD$. The partition function of the dimer model on
$\cD$, $Z_\cD$, which is related to that of the original Ising model
in a simple way, can in that case be evaluated by Pfaffian methods; by
Kasteleyn's theorem,\cite{kasteleyn} one can assign directions to each
bond of $\cD$ such that the product of orientations when traversing
any elementary plaquette in a clockwise direction is odd. These
orientations can be used to define a matrix, $A$, which has entries
$A\ij=+1 (-1)$ if bond $\left<ij\right>$ is oriented for $i$ to $j$
(from $j$ to $i$). With fugacities of the external and internal bonds
given by $w\ij$ and 1, respectively, one has $Z_\cD=\Pf A$.

A nonanalyticity in $Z$ bequeathes one to $Z_\cD$. For instance, the
Fisher lattice dimer model obtained from the square lattice 
ferromagnetic Ising model must exhibit a phase transition as 
the ratio of external to internal fugacities is varied. Nevertheless
the Ising character of the transition seems to have disappeared 
{\em en route}---there no longer is an Ising symmetry to break. 
In the following paragraphs, we will see that the dimer model does
not break any symmetries whatsoever. Instead the Ising transition has
turned into a confinement-deconfinement transition in which the
free energy to separate two test monomers is the appropriate 
diagnostic. The universality classes of the two transitions coincide,
as they must, by virtue of both of them mapping onto the theory of a 
single gapless Majorana Fermion.

As $\cD$ is periodic, one can find the partition function by Fourier
transformation; $Z=\Pf A=\sqrt{\det A}=
\sqrt{\prod_q\det \tilde{A}(q)}$, where $\tilde{A}(q)$ is the Fourier 
transform of $A$ at wavevector $q$. The specific free energy, $\cF$,
is given by 
\bea
-N\beta\cF=\ln Z_\cD=(1/2)\sum_q\ln(\det\tilde{A}(q))\ .
\eea
This expression does indeed reproduce Onsager's formula for the square
lattice Ising model for any coupling strength
$K$.\cite{onsager,FishIsing} In particular, nonanalyticities occur
when $\det\tilde{A}(q)=0$. With\cite{FishIsing}
\bea
\det\tilde{A}(q)=(1+w^2)^2-2w(w^2-1)(\cos q_x+\cos q_y), \nonumber
\eea this
happens only for $q_x=q_y=0$ and $w=w_c=1+\sqrt{2}$ so that
$K_c=\ln(1+\sqrt{2})/2$, as it should.

What is the nature of the phases on either side of $w_c$? For
$w\rightarrow\infty$, only one dimer configuration survives, namely
one in which all external bonds and the bond linking the two internal
sites are occupied. This configuration breaks no lattice symmetries,
and is in that sense not a crystal. 
In the opposite limit, $w=1$, all dimer configurations have equal
weight and their ensemble also respects all lattice symmetries.
Evidently the transition does not involve symmetry breaking.

Instead, the two phases differ in their response to the insertion of
a pair of test monomers (sites not part of a dimer). It is not
hard to see that the high temperature phase is confining.
If one places a monomer 
on one end of an external leg, the site on its other end has to pair
up with another site in its cluster (group of sites obtained from one
original site by decoration), which in turn leaves the partner of the
latter site unpaired. Two monomers placed a distance $L$ apart
therefore exact a cost in free energy proportional to the minimal
number of unoccupied external bonds, which is proportional to
$L$. This point is thus in  a confined phase.
Intuitively, the low temperature phase involves a true dimer fluid
which should therefore allow monomers to be separted with finite free
energy cost. 

\begin{figure}
\centerline{ \psfig{figure=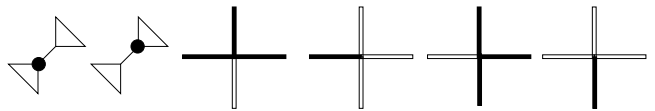,width=8.5cm} }
\caption{ 
The open string configurations (four rightmost plots) are generated
with a monomer (filled circle) placed on the upper interior 
point of the cluster. The other monomer location (leftmost plot) generates
the remaining four string configurations, obtained by a reflection of the
four displayed ones along a diagonal. }
\label{fig:monofish}
\end{figure}

These statements can be made precise by tracking the
spin-spin correlation, $\left<\sigma_i\sigma_j\right>$, 
from the spin formulation into the
dimer formulation.  The spin correlator can, in the context of the
high-temperature expansion, be expressed as a loop model on the square
lattice containing, along with the closed loops, 
one open string running between
the sites $i$\ and $j$.\cite{mpw} 
This loop-string model can in turn be cast in terms of a
dimer model with two monomers (Fig.~\ref{fig:monofish}).  The
vertices at the ends of the string can be encoded in the
monomer-dimer model by summing over four partition functions,
$Z_{md}$, with the monomers in the clusters $i,j$ 
being placed independently
onto one of two interior points of the cluster. 
The parition function of
the loop-string model is thus given by the sum over the four
monomer-dimer partition functions: 
$Z_{ls}(i,j)=\sum Z_{md}(i,j)$. 

Next, one uses $\left<\sigma_i\sigma_j\right>=Z_{ls}(i,j)/Z_{{\cal
D}}$.  It follows that in the high temperature (paramagnetic) phase,
where $\left<\sigma_i\sigma_j\right>$ vanishes exponentially, all the
monomer pairs are confined. Conversely, in the low temperature phase,
$\left<\sigma_i\sigma_j\right>$ decays to a constant, which implies
that at least one (and probably all) monomer pairs are deconfined.

Two comments on related issues are worth making. First, dimer
models generally exhibit topologically disconnected sectors
under local moves---a classical precursor of the notion of
topological order for quantum problems \cite{wen} which we
shall invoke below for quantum dimer models. On the Fisher 
lattice there are four such sectors on a torus corresponding to the 
various combinations of an even or odd number of dimers intersecting the two
non-trivial loops. The Fisher map represents the Ising partition
function as a sum over all four sectors of the dimer
model. Evidently, all Ising spin configurations in a finite system
are accessible from one another by local spin flips, and the
topological distinction in the dimer model is an artefact of the 
bookkeeping from the perspective of the Ising model.

Second, dimer problems are solved by lattice Majorana fermions. As
the unit cell of the Fisher lattice has 6 sites, microscopically
the dimer formulation leads to as many Majoranas. However, at 
the transition fugacity $w_c$, two eigenvalues of
$\tilde{A}(0)$ vanish. These two combine to form one single Majorana
Fermion, to yield the known critical theory as in the Ising model. 
We note that given the lore on fermion doubling,\cite{franmathsene}
it is somewhat surprising 
that we obtain only a single Majorana Fermion from our lattice problem.
While this has been commented on from a different persepctive 
before,\cite{fermdoub}
we note that in the dimer formulation this conclusion arises from the 
manifestly asymmetric nature of our lattice, no longer invariant 
under reflections along both $x$ and $y$ axes. This leads to lattice 
derivatives different from the ones normally encountered when discretising 
the continuum.

\section{Frustrated Ising models as dimer models with positive fugacities}
\subsection{The modified Fisher construction}
Fisher's construction is quite general. The high temperature expansion
generates a loop model on any interaction graph $\cG$, and the rules
for generating the clusters out of the vertices depend only on the
number of legs of each given vertex. 

One prehaps not so desirable feature of the Fisher mapping is 
that it leads to negative dimer fugacities in the case of frustrated
models: there, not all interactions can be chosen to be ferromagnetic,
and hence the $v\ij=1/w\ij=\tanh K\ij$ are negative for some of the
bonds.

Here, we present a way of addressing this problem via a modified
Fisher construction on the dual of a two-dimensional planar lattice.
It proceeds by mapping each Ising spin configuration onto a link
configuration, $\{\tau\}$, on the dual lattice.

A given link of the dual lattice is occupied ($\tau=1$) if and only if
the bond of the direct lattice it crosses is frustrated. Such a link
configuration has the property that the site of the dual lattice at
the centre of a frustrated plaquette has an odd number of occupied
links emanating from it; for an unfrustrated plaquette, this quantity
is even. Clearly a spin configuration and its Ising reversed 
counterpart map onto the same link configuration on the dual lattice.

\begin{figure}
\centerline{ \psfig{figure=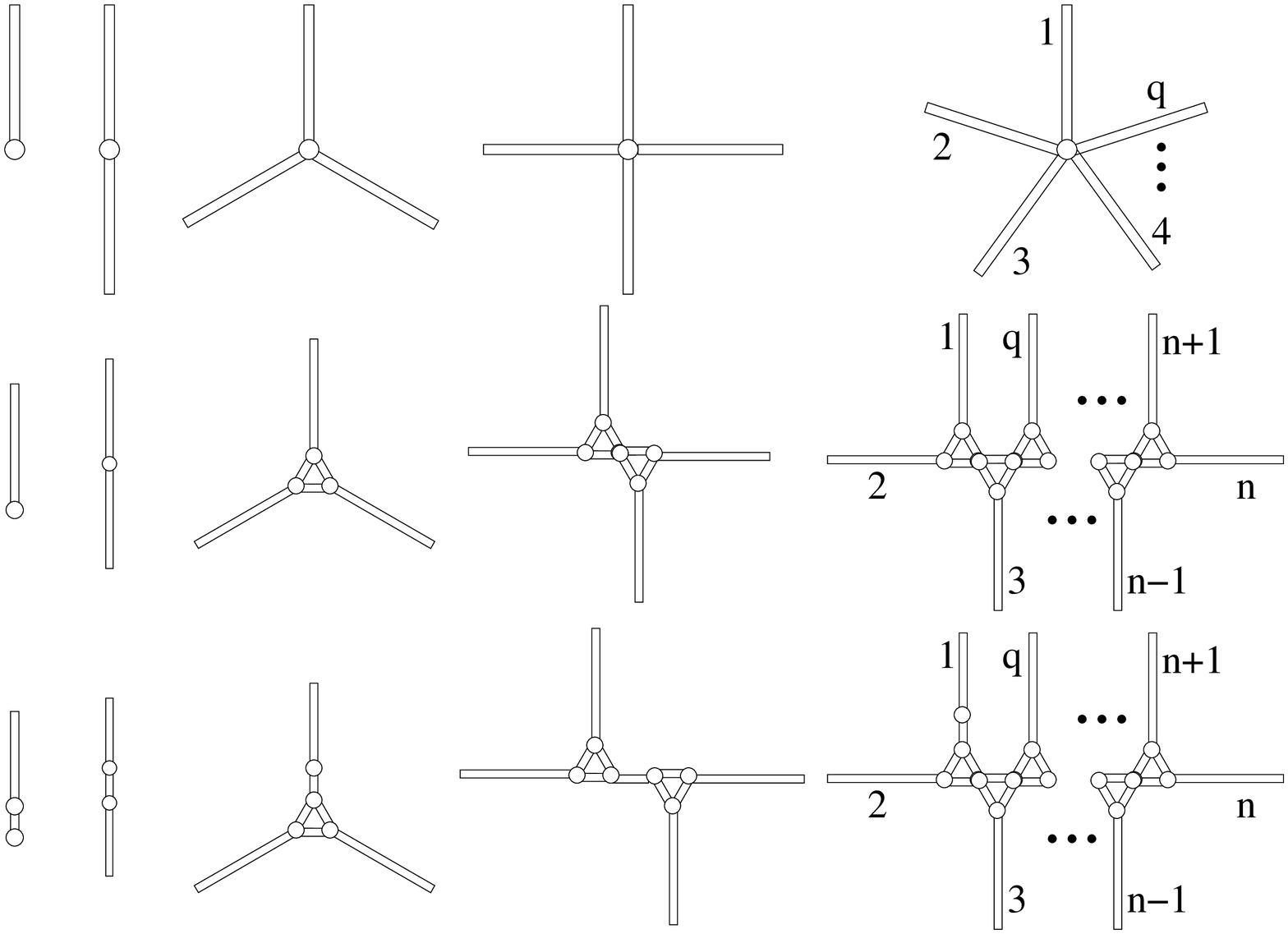,height=5.5cm} }
\caption{ 
Decoration of vertices of coordination $q$ (top) into clusters to
generate the lattice on which a frustrated (middle) or unfrustrated
(bottom) Ising model is represented by a hardcore dimer model with
positive fugacities. }
\label{fig:frcity}
\end{figure}

Such soft dimer configurations on the dual lattice can be converted into 
hardcore dimer coverings by suitably decorating each site.
Depending on the even or oddness of the number of occupied links at
the site, two different decoration operations
are required, leading to two different types of clusters. The decoration
transformation for odd sites is shown in Fig.~\ref{fig:frcity}. For
even sites, one requires a slightly different prescription as in
Fisher's original construction, as that included interchanging empty
and occupied links. The corresponding construction is also given in
Fig.~\ref{fig:frcity}.

The dimer fugacities are determined by the Boltzmann factors which
come with the presence of a frustrated bond. We find it convenient to
add a constant term to the Hamiltonian so that an unfrustrated bond
has energy $0$ and a frustrated bond has energy $2J$, so that the
dimer fugacity link $\left<ij\right>$ is given by 
$u\equiv\exp(-2K\ij)$,
which is always positive and between 0 and 1. As in Fisher's original
construction, Ising models on planar lattices lead to planar dimer
models, which can hence be solved using Kasteleyn's theorem.

Three comments on the differences between the Fisher construction
and the modified Fisher construction are in order.
First, as is appropriate for a dual construction, high and low temperatures
trade places when we compare the original Fisher construction to the 
modified construction---e.g. the equal fugacity dimer model corresponds
to zero and infinite temperature respectively. Second,  in
the modified construction the spin model maps 
onto a single topological sector of the dimer model. The remaining
sectors are generated by considering different boundary conditions
for the spin model. Third, the original mapping does not relate
individual spin configurations to dimer configurations but the
modified mapping does, upto a twofold ambiguity coming from global
Ising reversal. This will be important in making a connection between
Ising models and quantum dimer models below. But first we apply
the modified construction to the solution of a classical frustrated
Ising model.

\subsection{The fully-frustrated Ising model as a dimer model} 

As an an illustration of the above technique, we now use it to solve
Villain's odd model, also known as the fully frustrated Ising model on the 
square lattice.\cite{villainodd} This model is defined for spins on the
square lattice with nearest neighbour interactions such that each
plaquette has an odd number of antiferromagnetic interactions. This
can for example be achieved by choosing all horizontal bonds to be
ferromagnetic and the vertical bonds in ferromagnetic rows alternating
with antiferromagnetic ones. Transcribed to the dual lattice, which is
again square, this corresponds to an odd number of occupied links
emanating out of each plaquette. The corresponding decorated lattice
(Fig.~\ref{fig:fishlat}) has five sites per cluster.

\begin{figure}
\centerline{ \psfig{figure=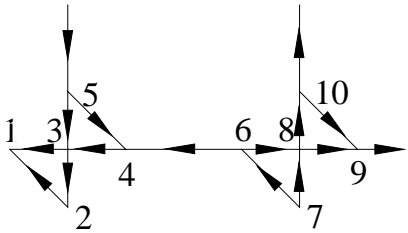,width=5.5cm} }
\caption{ Arrow convention for the Kasteleyn matrix of the
frustrated Fisher lattice.
}
\label{fig:frfisharrow}
\end{figure}

The Kasteleyn matrix on this lattice can be partially diagonalized
by means of a Fourier transform. While the lattice itself has five
sites per unit cell (see Fig.~\ref{fig:frfisharrow}), choosing 
appropriate signs for the matrix requires that we consider a doubled 
unit cell in one direction which we take to be the horizontal direction. 
This yields the $10\times10$ Fourier transformed matrix, $\tilde{A}(q)$, 
\begin{widetext}
\bea
\tilde{A}(q)&&=
\left(
  \matrix{ 0 & -1 & -1 & 0 & 0 & 0 & 0 & 0 & - {u}
      {e^{-\imag \,q_x}}   & 0 \cr 1 & 0 & -1 & 0 & {u}
    {e^{-\imag \,q_y}} & 0 & 0 & 0 & 0 & 0 \cr 1 & 1 & 0 & -1 & 
     -1 & 0 & 0 & 0 & 0 & 0 \cr 0 & 0 & 1 & 0 & -1 & 
     -u & 0 & 0 & 0 & 0 \cr 0 & - u e^{\imag \,q_y}\, 
       & 1 & 1 & 0 & 0 & 0 & 0 & 0 & 0 \cr 0 & 0 & 0 & u & 0 & 0 & 
     -1 & 1 & 0 & 0 \cr 0 & 0 & 0 & 0 & 0 & 1 & 0 & 1 & 0 & - {u}
      {e^{-\imag \,q_y}}   \cr 0 & 0 & 0 & 0 & 0 & -1 & 
     -1 & 0 & 1 & 1 \cr u e^{\imag \,q_x}\, & 0 & 0 & 0 & 0 & 0 & 0 & -1 & 0 & 
     -1 \cr 0 & 0 & 0 & 0 & 0 & 0 & u e^{\imag \,q_y}\, & -1 & 1 & 0 \cr  }
\right)
\eea
Its determinant, $\det\tilde{A}(q)$, is given by
\bea
\det\tilde{A}(q)&&=
  2\,u^2\,\left( 2\,{\left( 1 + u^2 \right) }^2 + 
     {\left( -1 + u^2 \right) }^2\,\cos (q_x) - 
     {\left( -1 + u^2 \right) }^2\,\cos (2\,q_y) \right) \ ,
\label{eq:detaq}
\eea 
\end{widetext}
which yields the dimensionless free energy per site, 
\bea
-\beta\cF=\frac{\ln u}{2}+\frac{1}{4}\iint_0^{2\pi}
\frac{dq_x\ dq_y}{(2\pi)^2}\ln\det\left(\frac{\tilde A(q)}{u^2}\right) .
\label{eq:freeenergy}
\eea
The first term is the ground state energy, as there is one
frustrated bond for each pair of sites. At zero temperature, the
second term gives the ground state entropy, which integrates to the
well-known result $G/\pi$, where $G$ is Catalan's
constant.\cite{kasteleyn,fisherdimer} 
After a little algebra, our expression
Eq.~\ref{eq:freeenergy} can be shown to agree with Villain's result 
for the partition function given in Appendix 2 of 
Ref.~\onlinecite{villainodd} at all temperatures.\cite{fn-compare}

A direct examination of Eq.~\ref{eq:freeenergy} shows that the model 
is critical only at $T=0$, where $u=0$. As a final excercise we will
now compute the divergence of the correlation length as 
$T \rightarrow 0$ by means of our solution.

To find its behaviour in the critical regime, we consider the case of
small $u$.  Near $u=0$, we can expand $\det({\tilde A(q)})/(2u^2)$ to
find that it varies as
\bea
2+\cos(q_x)-\cos(2q_y)+
2u^2(2-\cos(q_x)+\cos(2q_y))\ .
\label{eq:detqanrzero}
\eea 
In particular, at $u=0$, it vanishes at $p_1^{(0)}=(\pi,0)$ and
$p_2^{(0)}=(\pi,\pi)$.

To find the correlation length, it is necessary to compute the Green
function by inverting $A$. This is done by inverting $\tilde A(q)$ to
obtain and then carrying out the inverse Fourier transform on $\tilde
G(q)=\tilde A^{-1}(q)$. If one is only interested in the 
correlation length and not the details of the correlations, 
it suffices
to do the Fourier integral $\iint d^2q \exp(iqr) \tilde G(q)$
asymptotically, using the property that due to the inversion process,
the structure of $\tilde G(q)=\tilde g(q)/\det \tilde A(q)$, where
$\tilde g(q)$ denotes a cofactor. 
We thus
have to do integrals of the type
\bea
G\sim\iint d^2\vec{q} \exp(i\vec{q}\cdot\hat{r}R)/f(\vec{q})\ ,
\eea
where the zeroes of $\det \tilde A(q)$ determine the locations of the
poles of the integrand -- we have checked 
that the cofactors do not all vanish
at the locations of the poles. In this equation, we have emphasised the
two-dimensional nature of $q=(q_x,q_y)$ by writing it as $\vec{q}$ for
the time being. $\hat{r}=(\cos\theta,\sin\theta)$ 
is a unit vector in the direction of which
the correlations are to be computed, and $R\rightarrow\infty$ is the
quantity in which we will evaluate the integral asymptotically, so that
$(x,y)=R \hat{r}$.

Let us first carry out the integral over $q_x$: 
\bea
I_x\equiv\int
dq_x\exp(R i q_x\cos\theta)/f(\vec{q}). \eea
From
Eqs.~\ref{eq:detaq}, \ref{eq:detqanrzero}, we find that the location of
the poles can be written in terms of 
$\vec{q}^\prime=(2 q_x^\prime,q_y^\prime)$ so that
$\vec{p}_i=\vec{p}_i^{(0)}+\vec{q}^\prime$, where we have inserted a factor 
of 2 in front of $q_x^\prime$ to make the square symmetry more apparent:
\bea
{{\vec{q}^{\, \prime 2}}}
={q_x^\prime}^2+{q_y^\prime}^2=-4u^2\equiv-\alpha^2u^2\ ,
\eea
so that 
the poles lie
at 
$q_x^\prime=\pm\sqrt{\alpha^2u^2+{q_y^\prime}^2}$. 
Depending on the sign of
$\cos\theta$, we therefore choose an integration contour that runs
through the following points in the complex $q_x$ plane:
$(0,0)\rightarrow(2\pi,0)\rightarrow(2\pi,\pm\infty)\rightarrow(0,\pm
\infty)\rightarrow(0,0)$. Due to the periodicity of the integrand,
the contribution from the vertical contours cancel. The contribution
from the horizontal contour at $\pm I \infty$ vanishes for
$R\rightarrow\infty$ as $\exp(-R |\cos\theta|)$, so that $I_x$ only
picks up a contribution from the enclosed poles, where we denote
the residue as $1/f^r[\vec{p}_i(q_y^{\prime})]$.

To find $G$, we then need to determine 
\bea
G&\sim&\int_0^{2\pi} dq_y I_x\\
&=&
\int{dq_y^\prime}
\left\{
\frac{(-1)^x}{f^r[\vec{p}_1(q_y^{\prime})]}+\frac{(-1)^{x+y}}
{f^r[\vec{p}_2(q_y^{\prime})]}
\right\}
\times\nonumber\\
&&\times\exp\left[-R\left(
|\cos\theta|\sqrt{\alpha^2u^2+{q_y^\prime}^2}-iq_y^\prime\sin\theta
\right)
\right]\ .
\nonumber
\eea
This integral can be treated asymptotically using Laplace's
method. The dominant contribution comes from points where the argument
of the exponential is stationary: 
\bea
\frac{d}{dq_y^\prime}\left\{
|\cos\theta|\sqrt{\alpha^2u^2+{q_y^\prime}^2}-iq_y^\prime\sin\theta\right\}=0. 
\eea One
finds $q_y^\prime=i\alpha u\sin\theta$, so that at large distances, $G$
decays exponentially as $\exp(-R/\xi)$, where the correlation length
$\xi^{-1}=2u$.  As $T\rightarrow0$, the correlation length of the
Ising model therefore displays a divergence proportional to
$\exp(2J/k_BT)$, in agreement with the result 
given in 
Ref.~\onlinecite{zittcorr}, which was obtained using the transfer matrix
technique.

\section{The three dimensional Ising model as a two dimensional 
quantum dimer model} 

In this section we use the modified Fisher construction to
reformulate a class of three dimensional Ising models as quantum dimer
models in $2+1$ dimensions. The interest of this mapping is that it
yields non-trivial information on the quantum dimer models by
transcribing known lore on the Ising models---no advances in
solubility are entailed.\cite{fn-anydim}

The prototypical example of the Ising models of interest is the
nearest neighbour ferromagnetic Ising model on the cubic lattice.
As is well known, this is equivalent to a $2+1$ dimensional
transverse field Ising model in the ``time continuum'' limit.
Briefly, this proceeds by the recognition that the phase structure 
of the model is unchanged
if we take the cubic lattice to be a set of stacked square lattices
and allow unequal couplings in the plane and in the stacking
direction, 
$H (J_\tau,J_s) =-J_\tau \sum_{\{ij\}}\sigma_i^z
\sigma_{j}^z- J_s \sum_{\left<ij\right>}\sigma_i^z \sigma_j^z$
where the sums $\left\{ij\right\}$ and $\left<ij\right>$ run
over nearest neighbour sites in the stacking and planar
directions. 
The anisotropic scaling
limit $\exp(2 K^\tau)\rightarrow\infty$ can be identified with the
Trotter-Suzuki decomposition of the imaginary time path integral of
the two-dimensional transverse field Ising model
with the quantum Hamiltonian
$\hat{H}=-J_s \sum_{\left<ij\right>} \hat\sigma_i^z \hat\sigma_j^z -
\Gamma\sum_i\hat\sigma_i^x$. 

The Hilbert space of this quantum model is spanned by all classical
Ising configurations of the two dimensional classical model and the
transverse field moves the system between these configurations.
As the classical configurations can be related to dimer configurations
on the Fisher lattice by the modified Fisher construction, it should be 
intuitively clear that the model can equally well be cast as a resonance 
dynamics in the space of dimer configurations. 

\begin{figure}
\centerline{ \psfig{figure=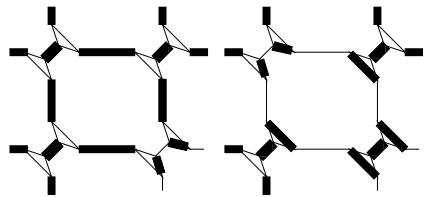,width=5.5cm} }
\caption{One kinetic energy term of the
quantum dimer Hamiltonian generated by the simple cubic Ising
ferromagnet has a matrix element between the two configuration shown here.
}
\label{fig:fishlatdyn}
\end{figure}

The quantum dynamics induced in this way can be visualised by noting
that the effect of the transverse field is to flip individual
spins. In the modified Fisher construction, this corresponds to
replacing empty {\em external} links surrounding the spin by occupied
ones, and vice versa. In the dimer model, this also entails moving
internal bonds to accomodate the change in external bonds. Here, it is
important to note that the configuration of internal bonds of a cluster,
given a set of external bonds, is unique and determined only by the by
the external bonds belonging to that cluster. The quantum dimer
Hamiltonian is therefore strictly local, although it includes a number
of kinetic energy terms and loop flips of varying length with exactly
the same strength---all nonzero 
off-diagonal matrix elements equal $-\Gamma$.
An example of a dimer move present in the current
Hamiltonian is given in Fig.~\ref{fig:fishlatdyn}. In addition
the Ising nearest neighbour interaction---the number of frustrated
bonds---translates into a potential energy for dimers on the
external bonds.

The resulting quantum dimer Hamiltonian of this model can be
schematically written as
\bea
H_D&\equiv&\hat V+\hat T\nonumber\\
&=&-\sum_{\left<ij\right>}J\ij\hat\tau^x\ij+
\Gamma\sum_\Box\prod_\Box\hat\tau\ij^\pm\ .
\eea
The $J\ij$ of internal bonds are zero and the sum $\sum_\Box$ runs
over all closed loops made up of the external bonds surrounding a spin
of the direct lattice and any of the bonds of the clusters on which
these bonds terminate. 
$\prod_\Box\hat\tau\ij^\pm$ stands for
alternating raising and lowering operators, $\cdots
\tau\ij^+\tau_{jk}^-\tau_{kl}^+\tau_{lm}^-\cdots$ as one goes around the
loop; this form preserves the hardcore condition.

In the above we have ignored one subtlety, namely that the map 
between Ising 
configurations and dimer configurations is two-to-one and the Ising
dynamics {\it does} connect a state and its Ising reversed counterpart. 
The solution to this lies in considering 
combinations of a given state and its reversed counterpart\cite{msc}
\beq
| \{ \sigma_i \} \rangle_{\rm e/o} = 
{1 \over \sqrt{2}} \left(
| \{ \sigma_i \} \rangle \pm | \{- \sigma_i \} \rangle
\right)
\eeq
that are even and odd under global Ising reversal---a property
respected by the Ising dynamics. In order
to fix the ``up'' states we choose those so that a particular
spin is always up in them. Now both sets of states can
be mapped onto dimer states but the dimer dynamics is slightly
(but importantly) different in the two sectors. In the even
sector it is exactly what we described above. In the odd
sector most matrix elements are the same but the ones that involve
the chosen spin acquire an extra minus sign. Since $\Gamma$
can be chosen negative without loss of generality, by the Perron-Frobenius
theorem the ground state will always be in the even sector,
although the first excited state need not be.
So for the purposes of determing phase structure one can ignore
this complication entirely. From the perspective of the dimer
model, which is what we will take in the remaining, its dynamics
will be represented solely by the even states of the Ising
model.

At zero temperature, there is one parameter in this problem, namely 
the ratio of transverse field to bond strength, $\Gamma/J_s$ and two
phases that meet at a critical point. 
In Ising language, the two phases are, of course, the ferromagnet and 
paramagnet. 
As the reader may anticipate from our previous
classical considerations, the dimer transition is between a
deconfining phase at large $\Gamma/J_s$ and a confining phase
at small $\Gamma/J_s$ neither of which break any lattice
symmetries.

\begin{figure}
\centerline{ \psfig{figure=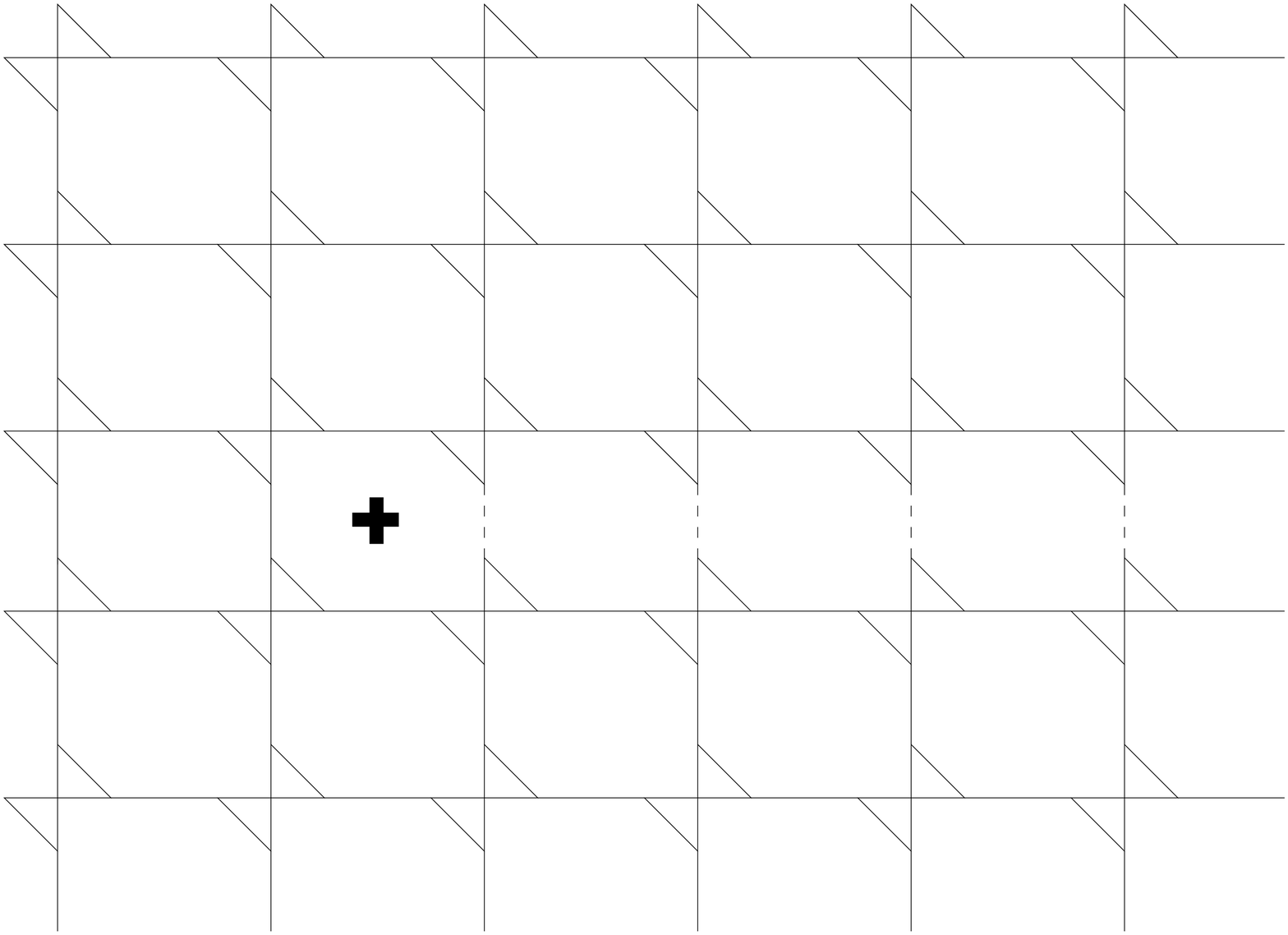,width=0.7 \columnwidth} }
\caption{
A spin antialigned with the transverse field at the centre of the
plaquette denoted  by the cross translates into a vortex excitation
in the quantum dimer model.}
\label{fig:fishlatvortex}
\end{figure}

The deconfined phase is of particular interest. Its point $\Gamma/J_s
= \infty$ is the ``Rokhsar-Kivelson'' point
of the model where the ground
state wavefunction is the equal amplitude sum over all dimer
configurations, which is the canonical short ranged RVB state. As
discussed in Ref.~\onlinecite{misserpas}, the elementary Ising
excitation is a a spin antialigned with the transverse field with
energy $2\Gamma$, and the entire spectrum is composed of them.  In the
dimer model the flipped spin translates into a vortex in which dimer
configurations pick up a minus sign if the number of dimers on a
string extending from the chosen plaquette to infinity is odd
(labelled by dotted bonds in Fig~\ref{fig:fishlatvortex}). The dimer
model is exactly equivalent to the standard Ising gauge theory, and
the vortex is therefore an Ising vortex (vison).  For a system on a
torus, these vortices need to be created in pairs, so that the minimal
excitation energy is in fact $4\Gamma$.\cite{misserpas}

The Ising states with a single flipped spin have disappeared in the
course of the duality transformation. In fact, together with the other
states with an odd number of flipped spins, they make up the sector,
odd under global Ising reversal, which was discarded {\em en route} to
the Ising model. The ground state at the Rokhsar-Kivelson 
point in this sector is
the first excited state of the Ising model and has a degeneracy
proportional to the system size. The equal amplitude superposition
of the configurations in that sector corresponds to the reference spin
being antialigned with the field.

As observed before there are four dimer sectors on the torus (and
$4^g$ on lattices of genus $g$) which cannot be connected by local
dimer moves. As these sectors correspond to different
boundary conditions in the Ising problem, and the latter have only an
$O(e^{-L/\xi})$ effect on the ground state energy in the paramagnetic
phase ($\xi$ is the correlation length), it follows that they lead to
four exponentially degenerate ground states. Hence the deconfined
phase displays topological order in the sense of Wen.\cite{wen}

Finally we observe that the dimer model on the Fisher lattice is
exactly equivalent to the standard Ising gauge theoy on the square
lattice -- dubbed the ``even'' Ising gauge theory in
Ref.~\onlinecite{msf} due to the nature of its constraint. That
constraint states, in the Hamiltonian formulation, that the number of
units of Ising electric flux entering/leaving a site (the distinction
is irrelevant for Ising variables) must be even. If we identify a
dimer on the external bonds of the Fisher lattice with such a flux, we
recover this constraint.  This conclusion is only to be expected since
the even Ising gauge theory is dual to the transverse field Ising
model. 

Interestingly, the dimer constraint on the Fisher lattice, that the
number of dimers coming out of a given site be one, is a $U(1)$
constraint. As discussed in Ref.~\onlinecite{msf}, this gives the
dimer model the character of a $U(1)$ gauge theory at the lattice
scale. As in the case of the triangular lattice QDM, the topological
sectors indicate a low energy structure with Ising character -- which
would be the general expectation for a deconfined (RVB) phase. What is
special here is that, unlike the triangular QDM, we are able to put
all states of the Fisher lattice QDM under consideration into
correspondence with states in the Ising gauge theory.

Our basic considerations in this section can be generalized to a wide
class of Ising models that permit a time continuum limit to be taken.
These include models that can be viewed as identical planes stacked
ferromagnetically or antiferromagnetically (or in any alternating 
combination). Depending on the frustration in the
planes, we will obtain a QDM on a specific lattice obtained by the
decoration procedure. For example, the FFIM stacked ferromagnetically
will give rise to a QDM on the lattice shown in Fig.~\ref{fig:fishlat},
which will be exactly equal to the ``odd'' Ising gauge theory in which
the number of Ising fluxes leaving a site is odd. This QDM will
exhibit a transition from a deconfined phase to a confining phase
that does break lattice symmetries, as first discussed in the language
of Ising models in Ref.~\onlinecite{blankschtein}.

\section{Comments on the soluble kagome quantum dimer model}

The dimer model we obtain at $\Gamma/J_s\rightarrow\infty$ has much
in common with the kagome dimer model discussed by Misguich
\etal: their quantum dimer model also has a range of kinetic terms around a
given hexagonal plaquette, and it displays an analogous excitation
spectrum consisting of Ising vortices. Nevertheless, the solubility
and beautiful simplicity
of their model does not derive from a Fisher construction.
Instead, their problem has another very useful ingredient, 
namely the mapping
of hardcore dimers to arrows on the kagome lattice,\cite{elserzeng}
and indeed any other lattice consisting of corner-sharing triangles.

Not until they have executed this mapping 
does our discussion of the Rokhsar-Kivelson
point parallel their model, as one 
can consider the arrows as {\em bond} variables of a triangular
lattice Ising model (or equivalently, link variables $\tau^x$ of a
honeycomb gauge theory) which, crucially, has zero exchange
strength. The quantum dynamics they study consists of the Wilson loop
operation $\prod_{\hexagon}\tau^z$.

\subsection{An exactly soluble quantum eight vertex model}

To illustrate this correspondence, and because it is of interest to
ask if the arrow mapping used by Misguich \etal\ can be generalised,
we note that their approach can be used to define an exactly soluble
quantum eight vertex model. This model can also be given an
interpretation in terms of singlet bonds.

The basic geometric object of this model are squares (instead of
triangles), which are again arranged to share corners.  At the corners
of the squares reside Ising degrees of freedom (arrows), which point
either in or out. In addition, we impose the constraint that an even
number of arrows point out.

\begin{figure}
\centerline{ \psfig{figure=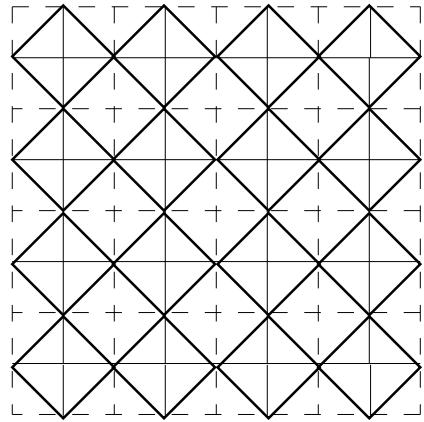,width=5.5cm} }
\caption{
A lattice of corner sharing squares (thick lines), which is the medial
lattice of the lattice denoted by thin lines, which is in turn
dual to the lattice denoted by dashed lines. }
\label{fig:eightarrow}
\end{figure}

The simplest case occurs when the squares (denoted by fat lines in
Fig.~\ref{fig:eightarrow}) are arranged on a square lattice (thin
lines) so that they share corners. The different arrow configurations
can be mapped onto configurations of the eight vertex model on the
dual square lattice by identifying the arrows (which by construction
live on the links of the dual lattice) with the arrows of the vertices
of the eight-vertex model.

One can further interpret the arrows as link variables of an Ising
model. The sites defining this model live on a further dual lattice
(dashed lines). For the case of a bipartite thin lattice, one can
label a bond frustrated if the arrow points from sublattice A to
sublattice B and unfrustrated otherwise. (In the case of a
non-bipartite thin lattice, one can choose a reference arrow and a
reference spin configuration and identify the two). The constraint of
the eight-vertex model then becomes a constraint on the product of
exchanges, $\prod_{\Box}J\ij=1$, around a plaquette of the dashed
lattice. However, this only imposes a constraint on allowed states --
the strength of the exchanges vanishes. From this method, one can
easily visualise the known result that the number of eight-vertex
configurations on the square lattice equals $2^N$, as there are no
constraints on the Ising model on the dashed lattice. This result is
straightforwardly generalised to any lattice defined by the midpoints
of corner-sharing squares, as explained in Ref.~\onlinecite{misserpas} 
for 
lattices defined by midpoints of corner-sharing triangles.

This eight-vertex model can again be endowed with a quantum dynamics
generated by a Wilson loop action on the links of the elementary
plaquette of the thin square lattice. This becomes the zero-exchange
transverse field Ising model on the dashed square 
lattice.\cite{fn-sixvertex} 
The
results, such as on the gap, $4\Gamma$, or on the ultra-short
correlation, follow just as they did before.\cite{misserpas}

\begin{figure}
\centerline{ \psfig{figure=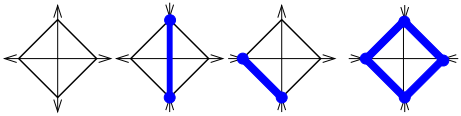,width=0.7\columnwidth} }
\caption{Possible interpretation of the eight vertex configurations
(denoted by arrows on thin lines) in terms of singlet bonds (denoted
by dimers) between spins 1/2 on the corners of the square.  The
rightmost figure denotes an overall singlet of the four spins, of
which there are two linearly independent ones.  }
\label{fig:eightvertexconf}
\end{figure}

A possible interpretation of the different eight-vertex configurations
in terms of singlet bonds is as follows (see
Fig.~\ref{fig:eightvertexconf}). Let spins $S=1/2$ reside on the
vertices of the fat square lattice. If no arrows point into a given
square, there are no singlet bonds between any pair of spins of the
square. If two point in, there is a singlet bond between the two. If
four point in, the spins on the square form some collective singlet
state. The choice of equal fugacities for all vertices amounts to
disregarding the entropic contribution of the two linearly independent
collective singlets.

As a related (and known) combinatorial result, we notice in passing
that the number of non-overlapping (but possibly intersecting) loop
configuration on a lattice can be trivially evaluated in the same
spirit. One can think of the loops as domain walls of an Ising model
on the dual lattice. Up to a factor 2 from the global Ising symmetry,
the number of domain wall configurations is then given by the size of
the configuration space of the Ising model, which is of size $2^N$ for
a dual lattice with $N$ sites.

\section{Summary}
Two dimensional Ising models are intimately related to dimer models.
By means of the modified Fisher construction introduced in this paper,
this connection can be exhibited configuration by configuration.
The Ising transition in two dimensions maps onto a deconfinement
transition in the dimer language as does the Ising transition in
three dimensions which now appears in a quantum dimer model. This
last equivalence provides an instructive example of a topologically
ordered RVB phase and a transition from it to a confining phase,
and one for which all the details are known. These ideas can be 
extended straightforwardly to generate an exactly soluble quantum 
eight vertex model.

\section*{Acknowledgements}
We would like to thank Gregoire Misguich, Vincent Pasquier and Didina
Serban for useful discussions, and Paul Fendley also for collaboration
on related work.


\begin{thebibliography}{99}
\bibitem{fn-compare}
In Villain's expression, there is a different choice of ground state
energy and the bond strength he uses is such that
$2J_{Villain}=J$.
\end{thebibliography}
\end{document}